# TRENDS TOWARD REAL-TIME NETWORK DATA STEGANOGRAPHY

James Collins, Sos Agaian

Department of Electrical and Computer Engineering
The University of Texas at San Antonio, San Antonio, Texas, USA
james.collins@utsa.edu, sos.agaian@utsa.edu

*Abstract*

*Network steganography has been a well-known covert data channeling method for over three decades. The basic set of techniques and implementation tools have not changed significantly since their introduction in the early 1980's. In this paper, we review the predominant methods of classical network steganography, describing the detailed operations and resultant challenges involved in embedding data in the network transport domain. We also consider the various cyber threat vectors of network steganography and point out the major differences between classical network steganography and the widely known end-point multimedia embedding techniques, which focus exclusively on static data modification for data hiding. We then challenge the security community by introducing an entirely new network data hiding methodology, which we refer to as real-time network data steganography. Finally, we provide the groundwork for this fundamental change of covert network data embedding by introducing a system-level implementation for real-time network data operations that will open the path for even further advances in computer network security.*

## 1. INTRODUCTION

Even though the origins of steganography reach back to the time of ancient Greece, to Herodotus in 440 BC, the art of steganography continues to evolve. This is especially true in the technical methods of digital embedding [1][2]. Digital steganography has been used since the early 1980's and network steganography techniques quickly evolved with the advent of the Internet and standardized multimedia formats used for data exchange [3]. These early network steganography methods have remained consistent in technique with only minor variances associated with network applications. We will introduce the next logical evolution of network steganography; moving from the simple static data modification class of attacks on data-at-rest, to a more complex set of threats that involve modification on the transition operations involving data-in-motion and data-in-use [4].

Figure 1 shows the interaction of the cyber threat domain interfacing with both the data-at-rest and data-in-motion domains. Data-at-rest steganography is characterized by embedding applications that target static multimedia files. Data-in-motion embedding methods enters into the domain of network protocol modifications. This interaction defines the concept of network steganography. Network steganography is essentially the exploitation of network elements and protocols for implementing covert communications [5]. These threats imposed by network steganography are generally considered one of the most difficult challenges facing private and government organizations due to not only the technical complexity and sheer volume of data analytics involved but also from the national and international law and politics [6][7].

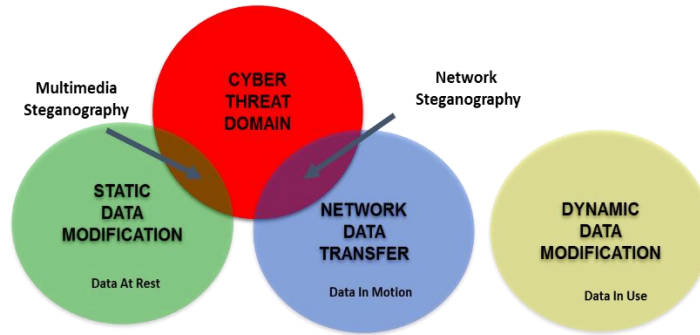

Figure 1. Cyber Threat Interfaces

There are currently three major accepted classifications of network steganography which are the protocol storage, protocol timing, and application protocol header modifications [3][5]. Each of these methods takes advantage of the open system TCP/IP network communications protocols to exploit and establish covert data channels. Some methods are more efficient and robust than others and key features for each will be detailed in this paper.

## 2. NETWORK STEGANOGRAPHY CONCEPTS

The idea of using the underlying protocols or elements of the fundamental operation of the networks dates by to the early 1980's during the formation of the early Internet and the defining period for many network protocol definitions [3][6]. In fact, a review of several key online databases shows that before the year 2000 only 25 papers discuss the topic of network steganography. Since that year, to date, over 1000 papers have been cited and more than a dozen sophisticated network steganography tools have been developed and released as open source [8]. Table 1 show a list of the most used or cited tools with a brief description on each [8]. These tools represents another avenue for cyber domain exploitation, such as the covert delivery or exfiltration of data, using the network protocol itself. Network steganography tools are directly contrasted from the many applications used for static data embedding. More specifically, these tools do not rely on the embedding of data into carrier files, as is the case with traditional multimedia but instead focus on the actual network protocols [6].

Table 1. List of Common Network Steganography Tools

| *Tool* | *Year* | *Functionality* |
|---|---|---|
| CovertTCP | 1997 | Covert Channels using TCP and IP headers |
| StegTunnel | 2003 | Timing channel using TCP header fields |
| hCovert | 2005 | Covert channel using HTTP GET requests between webservers |
| VoVoIP | 2007 | Embedding data in PCM voice traffic exchanges |
| SteganRTP | 2007 | Uses RTP of VoIP as payload medium |
| Gary-World Team's | 2008 | Covert channel projects using TCP and IP headers |
| Steganography Studio | 2009 | Training suite on network steganography tradecraft |
| NetCross | 2010 | Utilizes the DNS protocol to establish covert comm. |
| OpenPuff | 2011 | Multiprotocol embedding toolkit |
| SoCat | 2013 | Network relay transfer between two independent data channels. |

## 2.1 STEGANOGRAPHY MODELS

The seminal model used to describe covert channels in the field of steganography and steganalysis is known as the prisoner problem [9]. In this classic model, depicted in Figure 2, supposes prisoners are attempting to exchange information while keeping it hidden from a watching warden. The prisoners can hide the messages into the cover file and use a prearranged shared key as well as a randomization variable to obfuscate the data [9][10]. Because the warden does not know if the data exchange contains hidden information, the warden can act in three ways; passive, active, or malicious. In the passive warden mode, the messages can be observed but not modified. With active mode, the messages can be slightly modified without any obvious distortion or disruptive effects noticed by the exchanging prisoners. Finally, in active mode, the messages can be completely modified even to the point of altering the semantic content or introducing distortive effects [9][11].

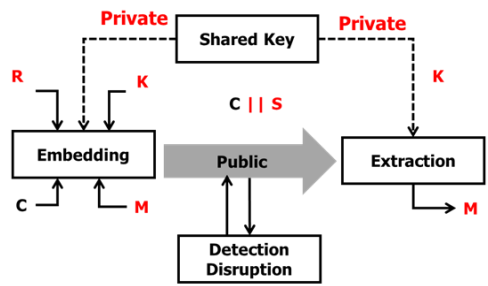

Figure 2. Classic Prisoner Problem

Network steganography is built upon this same prisoner-problem model [3]. The only major difference is that medium or cover file used is the network. When employed specifically on the TCP/IP protocol-based Internet, is the network protocol itself and not a traditional carrier file (e.g. image, audio). The standard internetworking model is known to be based on a client-server architecture consisting of several protocol regions of operation. Figure 3 elaborates on this model to show three distinct segments of interest. Region 1 consist of application level data. Region 2 is comprised of the structures and functions involved in internetworking and protocol exchanges. Region 3 involves the lower level physical data link control operations for moving data on to, off of, and between network components; for example routers and switches.

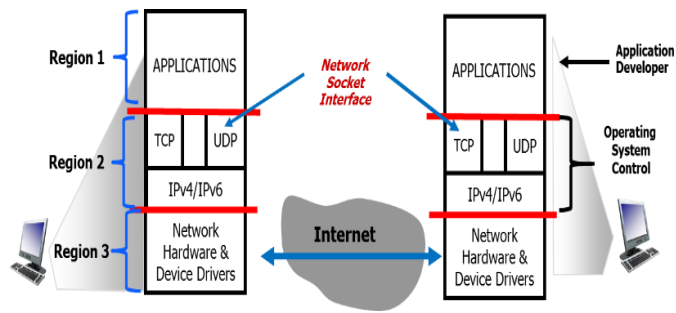

Figure 3. Internetworking Model of Network Steganography

Using this model and relating it back to the prisoner-problem, it is noted that the network steganography practitioner has full access and control to portions of the application, TCP or UDP, and the IP headers to implement a protocol centric covert channel exchange over. More specifically, the prisoners control all aspects of Region 1 and Region 2 operations since they are integral to the end system operations. Figure 4 shows the span of data control ownership at the end-point system, or that which the prisoner controls. This figure clearly shows that in the client-server paradigm of the Internet, a majority of the command and control of the data exchange is influenced by the end user [12][13]. It is this control structure, which allows network steganography to be implemented between communicating end systems on the Internet [14].

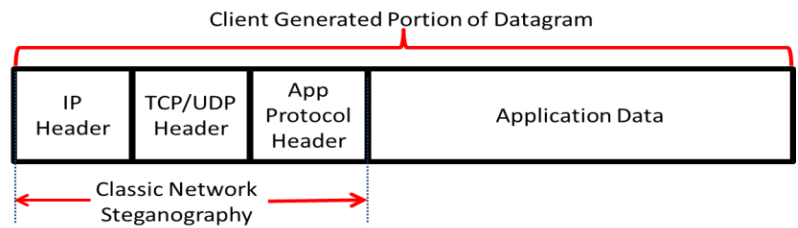

Figure 4. Region 1 and Region 2 Generalized Data Structure

Back to Figure 3, the Region 3 data will typically be structured as shown in Figure 5 below. In this case, the underlying network and device drivers use the Ethernet datalink protocol to encapsulate and transport the upper level data. In classic network steganography, Region 3 data modification is not consider and is not affected by Region 1 and Region 2 modifications. In real-time network data steganography the entire span of the protocol structure will be considered which we describe later in more detail.

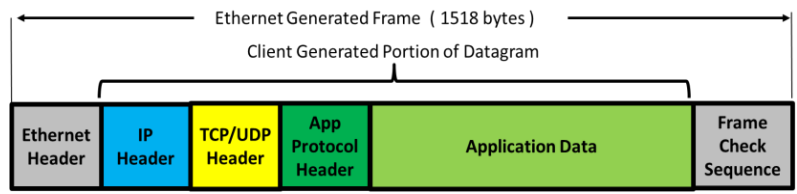

Figure 5. Typical Region 3 Data Structure

## 2.2 CLASSIC NETWORK STEGANOGRAPHY CHANNELS

In order to properly convey this diverse set of information involved in TCP/IP network application-based steganography, the generalized concepts of covert channels will first be used to describe two major classifications models. These are the *storage channels* and *timing channels* shown in Region 2 in the previous Figure 3 [3]. A third method, which has gained some increased recognition over the last several year, is referred to as the *application protocol channels* [3].

First, we define the overarching principle characteristics of these covert channels. In the first major category, storage channels are essentially employed by manipulating prescribed elements in data structures from which the sender and receiver can extract the prearranged coded values [15]. The second type of network steganography are the timing channels. Timing channel steganography is defined where covert communications is established by the artful modulation of a shared resource over a prescribed time period in order to effect an information exchange [16]. The third type, application protocol channeling, the application-level protocol exchanges themselves are used to establish a covert communication channels [3][17][18]. These application protocol header exchange methods implement secret communications by using resources residing above the socket layer of the Internet protocol suite [19]. The next several sections continue with the descriptions for these three types of covert network channeling methods and detail some specific examples for each.

**2.2.1 PROTOCOL STORAGE CHANNELS**

Storage channel steganography is implemented by manipulating and interpreting information beyond what is normally expected within predefined fields of a data structures. The TCP and IP data structures used in digital networking have been found to be very susceptible to covert channel exploitation in a number of ways [20][21][22]. Figure 6 below details the structure for the IP datagram and TCP segment, which are in use today on the Internet [12]. The IP data structure shown is known as version 4. While a version 6 is in fact in wide use on the Internet, for brevity, this research investigation will focus solely on version 4 as this is the predominant protocol in used on the Internet making up over 98% of all network traffic [23].

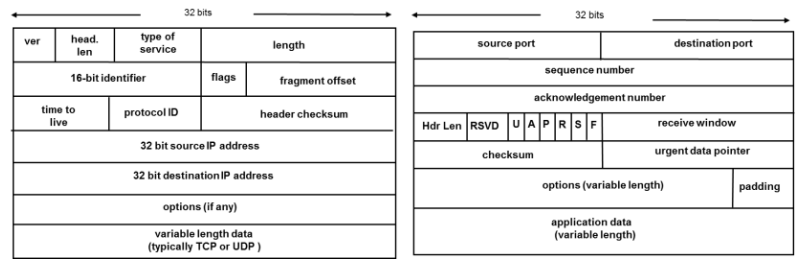

Figure 6. IP and TCP Data Structures

These two network protocol data structures and their field interpretations form the basis of a number of opportunities for network protocol steganography. Consider for instance the IP packet structure which consists of 13 specified fields. A number of the fields of the IP packet are known as mutable or changeable fields [13]. The mutable fields of interest are generally the identification, flags, fragment offset and time to live (TTL) fields. While these fields are subject to modification during transmission through the routers on the Internet, they can still be used as covert channels from sender to receiver [3][6]. A judicious modification to data in these fields does not affect the normal operation of the transfer protocol and so any well fashioned cover data often goes unnoticed and undisrupted [3][21].

One well known example of using these IP mutable fields as a covert storage channel is in the use of the identification, flags, and fragment offset fields [3][6][20]. In normal network datagrams transfers, these fields are used to control the fragmentation and reassembly of oversized datagrams that might pass through routers with datagram size restrictions. The datagram size for a data exchange can be determined by sending probing packets that discover the maximum transmission unit (MTU). The steganographer can discover the average, smallest MTU during a discovery probe and then simply choose a size even lower than this. In this way they are almost guaranteed that datagrams will not be segmented. Since the client controls the forming of the datagrams with the IP head, these fields – identification, flags, and offset – can be used to send embedded data of their choosing yielding up to 32 bits of covert data per packet header [20]. A similar method can be used to transmit data using the TTL field [24][25]. Since the TTL field is decremented by one at each router along the transmission path, a covert channel can be established that uses the upper two, three, or even four bits of this eight-bit field. The range of bit usage in this technique is clearly dependent on the number of intermediate routers. The number of path routers can however be discovered though a similar discovery probing method, namely *traceroute*. Other fields in the IP protocol header have been used for network steganography such as the type of service, protocol, header check sum, and the options field with boundary padding [3][24][26]. All of these IP header storage channel techniques have been successfully demonstrated on the Internet, each with varying levels of persistent success and overall covert channel bandwidth[3][24][25][26].

The TCP protocol header has also been used for network storage channel based steganography [3][16]. Unlike the IP header fields which can be modified by intermediate routers over the mutable field variables, the TCP header remains untouched and immutable by the datagram routing process over the Internet [12][3]. The TCP segment is completely managed and modified by the protocol stack software operating between two end-to-end client systems, thereby giving the steganographer full access to the data structure as a storage medium[13][14].

The most well know TCP based steganography storage channel is related to the sequence number fields [24][27][28]. To understand how this might be possible, consider that in a normal data exchange, the sender will generate an initial sequence number from which the sender and receiver then track and exchange the current state and count of the number of data bytes transferred. Given that the sequence fields are 32 bits in length, the most significant bits of this field could be used to send covert data for each TCP session established. This is just one simple of how the TCP header could be used and there are a number of more complex covert channel implementations that us TCP sequence exchanges. And just like with the IP header, the other fields within the TCP data structure – window sized, checksum, urgent point, options, padding have been used as covert storage channels as well [24][27][28].

General research in storage channels for network protocols tend to center on the TCP and IP protocol data structures. It should be noted that while these are the predominant protocols, their use is not exclusive. Referencing Figure 3 again, one may note that in Region 2 of this protocol model there is the user datagram protocol (UDP) structure. UDP is a very common protocol on the Internet used in extensively in gaming, video, and audio streaming applications. A number of steganographic network storage channel methods have been described as using the UDP transport [24][29][30]. It is also worth noting that as shown in Region 2 of the protocol model a space exists between the TCP and UDP descriptors. This is due to the fact that proprietary or user unique applications can be developed and implemented by clients and/or servers on the Internet. The open system architecture model embraced by the Internet paradigm allows for a plethora of Region 2 protocols to exist. This fact is what often drives the notion that the Internet communications protocol is virtually riddled with covert channels [26][31][32].

**2.2.2 PROTOCOL TIMING CHANNELS**

Timing channels can be implemented by modulating the packet transmission rate, changing the sequence timing, or modifying the packet loss or packet order [3][24]. Perhaps the best way to visualize the implementation of this type of network steganography is to view the timing diagram of Figure 7 below. This figure shows the sequence of packet transmission and arrival over time period of a client-server data exchange. This exchange is fully managed by the TCP software module on each of the end systems during the data exchange sequence and represents the case for end-user covert channel management. The TCP fields of interest during this transaction can include the sequence, acknowledgement, flags, windows size, and urgent pointer field [12][13].

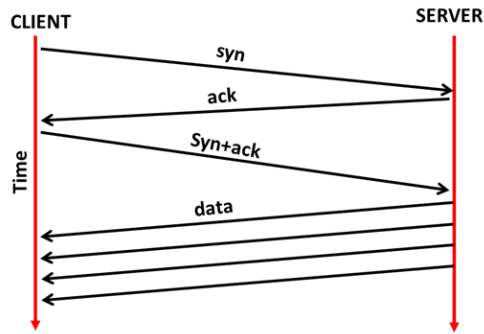

Figure 7. TCP Data Timing Exchange

In addition to these discrete protocol fields, there are associated timers that are set and managed by each of the end systems. It is the manipulation of these timers, along with the field values that can result in somewhat sophisticated and complex series of cover timing channels. The implicit modifications in the timing or sequence of events are what can establish uniquely interpreted changes that can result in covert time-based data exchanges between two participating end user systems [24][33].

As an example implementation of timing channel network steganography consider the previous figure that shows four datagrams being delivered in succession. The timing for the delivery of these packets is managed by special timers in the TCP software on the end user systems [12][13]. By slightly modifying the time for transmission, perhaps by +/-100 ms, the client and server can easily exchange a controlled series of binary values [24]. Another implementation involves the transmission and subsequent delivery of out of sequence packets [33][34]. Again, this is controlled by the end systems so carefully modulating the order of delivery can represent a unique series in a time-based binary stream [24].

### 2.2.3 APPLICATION PROTOCOL CHANNELS

In Figures 5 we saw the defined protocol headers for managing the internet protocol, the transport protocol (consisting of a TCP or UDP header) and also for the application protocol. It is worth noting that while the internetwork and transport headers are required elements of the protocol exchange and routing process, the application header is not mandatory and is often not used. In Figure 6, we also observe a well-defined canonical format for the two defined network protocol headers. When we consider the application protocol header, this is not the case.

The application header is used by the end systems to interpret and otherwise manage how the actual application data will be handled. Given that countless applications that could be developed, no single standard can be defined. Moreover, the Region 1 area of the protocol stack, as shown in Figure 3, is controlled 100% by the end user systems. We know that this is the same region in which, for example, multimedia data embedding can occur.

At this point, we highlight the fact that there is a distinct difference between end point application data embedding and application-protocol steganography channels. With end-point originated application steganography, the data is modified independent of the application protocol header [15][24]. With application-protocol steganography, any given protocol application can be used for a covert channel, but will remain fully independent of the data being managed by the actual application [29][35].

There are a number of published research studies that describe the means and methods to implement application-protocol covert channels [36][37][38]. Some specific Region 1 protocols investigated for exploitation have included HTTP, DNS, SMTP, FTP, and streaming applications over UDP with RTP and VoIP. We now briefly review a few of these common methods in an effort to further clarify the distinction between traditional multimedia application based steganography and application-protocol network steganography which involves the modification of the actual application protocol header exchanges.

**2.2.3.1 HTTP APPLICATION-LEVEL STEGANOGRAPHY**

Consider as an example, an HTTP based exchange. The HTTP protocol is a fairly well defined standard used to exchange information between webservers and web clients. There may be a number of different web client applications that could interface to a given webserver. For example, a webserver running Apache server code could interact with a number of different web browser such as Microsoft Explorer, Google Chrome, Firefox, Safari, and others. The end-node operating system could be Windows, Linux, Apple, Android, or a number of other obscure operating systems [39]. In order for the browser to be able to function properly with the webserver, an application level control header is used in the application region of the datagram, Region 1, to properly manage the overall data exchange. These transactions occur independently of the overarching transport control for functionality being performed by the TCP session exchange handlers [14].

A TCP session exchange can, for instance, operate normally without any timing or storage channel involved. Over this same session, the HTTP application protocol can be used as either a storage or timing delivery channel. This can be implemented independent of TCP by using, for example, the HTTP *close* or *persist* commands that would modify the sequence/time of delivery of the data [14]. An HTTP storage channel, again independent of protocol process Region 2, could be implemented using a number of mutable fields defined in the HTTP such as the structures surrounding the *put*, *post* and *get* commands [14]. Figure 8 graphically depicts the most basic protocol exchanges for several widely used internetworking applications. The basic process just described applies to these and countless other internetworking applications.

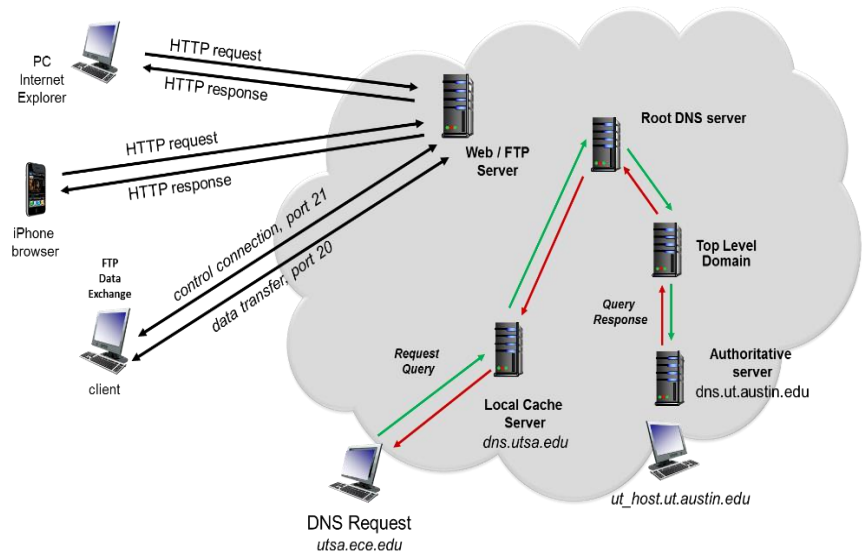

Figure 8. Application Protocol Exchanges

### 2.2.3.2 VOIP APPLICATION-LEVEL STEGANOGRAPHY

The rapid rise of transmitting voice data over the Internet using voice over IP (VoIP) followed by the development and subsequent explosive use of the VoIP application, *Skype,* spurred much research into how to use these environments for covert communications [29][30]. VoIP is a stream-based communications protocol that uses the real-time protocol (RTP) to set up and control the data flow [29][30]. An associated protocol, the real-time control protocol (RTCP) is used in conjunction with RTP to provide out-of-band statistics and control information for RTP sessions [12][13]. The basic structure of the RTP header is shown below in Figure 9 [12]. One described method of using this application protocol header for steganography is to merge the RTCP and RTP control information and embed unused bits in the IP, UDP, and RTP headers to signal various parameters that enable watermarking data in the voice stream. Another method, known as *SkyDe*, uses silence periods of a voice call to embedded encrypted steganography data and use the RTP and UDP headers to signal which packets are covert information [29][30].

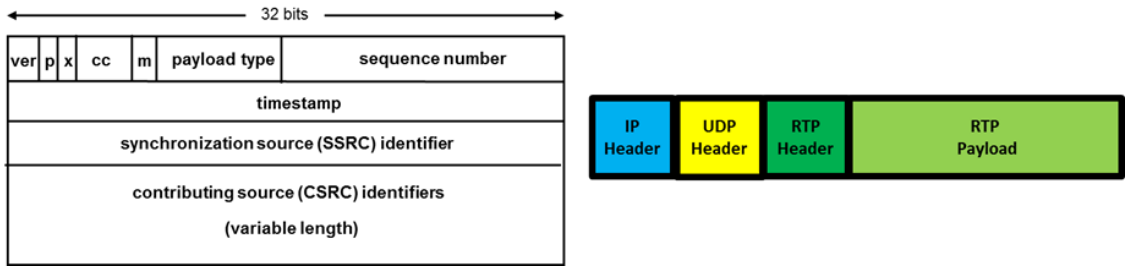

Figure 9. RTP Encapsulation and RTC Protocol Header

Each of these examples show that, like the more well-known storage-protocol and timing-protocols of IP and TCP, a much broader range of application-protocol header fields can be investigated for their potential to also be used to establish covert communications. These illustrative techniques form the major class of methods of what is known to be classical network steganography. The first two, storage and timing, are implemented strictly within the Region 2 space of the protocol stack. This latest method class, application-protocol, uses the Region 1 application header fields but may rely on some aspects of Region 2 to aid in the employment.

### 2.2.4 CLASSICAL NETWORK STEGANOGRAPHY CAPACITY

Multimedia files are used in traditional steganography due to the high capacity and as ubiquitous nature of the formats. The capacity of multimedia files such as images or audio formats is measured in terms directly related to the carrier file. For example an image of size 512x512 bytes, using a simple LSB embedding algorithm has an overall steganographic data capacity of just under 33,000 bytes. A simple audio wave encoding algorithm running against a file with 44100 samples per second has an embedding capacity of over 650,000 bytes in just 30 seconds. Network steganography, including all the classical methods described in the previous section, will have embedding capacities that pale in comparison with traditional multimedia steganography [1][2][3][24]. Covert data channels which use network protocol signaling have capacities which are measured in terms of data throughput using bits per packet, or sometimes bytes per second [3]. Before detailing some of the typical values for classical network steganography, some governing principles of network operation that constrain the channel bandwidth are presented.

## 2.2.5 NETWORK THROUGHPUT CONSTRAINTS

In referring back Figure 7, it was shown that the TCP data exchange between two communication points implements an initiation synchronization and acknowledge model [14]. The simplified transaction shown in this figure depicts an initial client-server contact followed by a data request and data delivery phase. The classical network steganography methods discussed in the previous sections are dependent on this fundamental communications transaction model, given that these defined covert channels are relying on the TCP and IP protocol header structures to convey the hidden information [24]. As such, the maximum achievable covert channel bandwidth or throughput for these classical network steganography methods will also be constrained by this model.

Figure 10 shows a more detailed time-based sequence of data exchange between a sender and receiver [14]. Several factors are evident from this diagram that contribute to the calculation of throughput for network steganography. First, this is a measurable transmission propagation time period, $T_p$, from the sender to the receiver. This time period is usually calculated in either milliseconds for network-to-network connections or in microseconds if the sender and receive are on the same network (i.e. LAN segment). Secondly, there is a slight delay from the receipt of the original message to the transmission of the acknowledgement message. This is the nodal processing time, $T_{np}$ that is typically measured in microseconds or nanoseconds [14]. Next, there is a return propagation timeframe once the receiver transmits the acknowledgement datagram. The entirety of this time period, from the beginning of initial sequence, to the receipt of the acknowledgement datagram is calculated as the network round trip time, *RTT,* for a TCP connection [14].

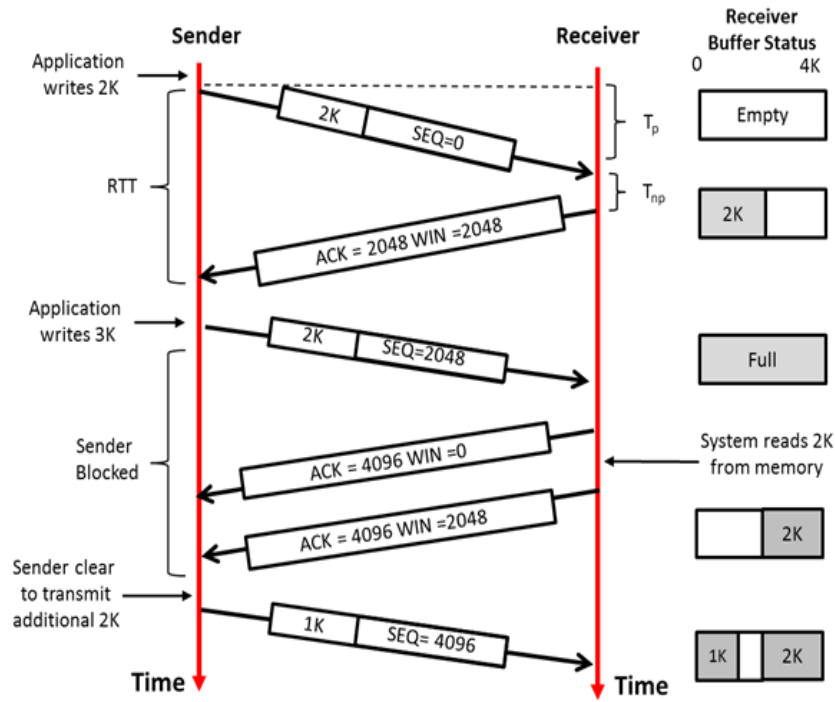

Figure 10. TCP Data Transfer with 4K Buffer

Any loss of data packets or TCP segments will ultimately affect the overall network protocol channel throughput or usability of the network for covert communications. Additionally, difficulty in manipulating and controlling TCP segment delivery will affect network steganography throughput [24]. These issues are all related to network congestion detection and control [14].

Referring again to Figure 10, another aspect of network data flow control that will come into play during classical network steganography is the management of how the applications read and write to a discrete defined buffer space during data transactions [24]. In the example shown in the current figure, a 4KB receiver memory space is reserved for packet data processing. Before the transmission of new data can occur, the sender must receive an acknowledgment that the receiver buffer is available. Buffer management will affect network timing when using TCP timing channels [3][14]. This same buffer configuration could also be used to actually implement a covert channel itself; such as sending or not sending data when the buffer is full or not full. In any case, buffer management becomes an important component to classical network steganography [3][24].

As an example of throughput comparison between traditional multimedia carrier-based data hiding and network steganography consider the following. With a standard BMP color image file of size 512x512 pixels, using only one color layer, the available carrier bytes is 262,144. This equates to an embedding capacity of 32,768 bytes if using simple LSB data embedding. Higher covert channel capacities are certainly achievable. Given this same image, which is 512x512x3 bytes, or approximately 786,432 bytes transferred over a TCP/IP network, a typical number of transactions from sender to receiver would be approximately 540 unilateral datagrams of size 1460. On a local area network this transfer would occur in a fraction of a second (e.g. 600 msec). If we consider using just the IP storage channel as previously described, we could potentially use 8 *bits* per packet. At best, this could achieve 540 bytes of transfer. Comparing 540 bytes to the approximately 33KB for traditional steganography adequately describes the contrast.

It is clear that the bandwidth of the classical network steganography channels are not as expansive as traditional multimedia carrier files. However, in one study of a specific network, it was estimated that each packet could be used to carry 8 *bytes* of covert data [37]. The test network cited externally transmitted approximately 500 million packets per day [37]. A malicious insider, able to exploit this steganographic network channel, could potentially exfiltrate data at a rate exceeding 4 GB daily, or approximately 1.5 TB annually – all via this often overlooked and unmonitored data channel.

## 3. NEXT GENERATION NETWORK STEGANOGRAPHY

Having described the fundamental background information for the various types of network steganography, in this section we develop the necessary background that will lead to the definition for *real-time network data steganography*. First, consider the Venn diagram as shown in Figure 11. The steganography cyber threat presented earlier is now expanded to intersect with all three data states, resulting in five distinct cyber threat subdomains. Two additional transitional subdomains are also shown; the data-at-rest to data-in-motion (DAR-DIM) region, and the data-in-motion to data-in-use (DIM-DIU) region. We assert that these two specific areas are concerned with normal system operations and data flow. These regions are fairly standardized interfaces which support application-level processing and network operations. We will however, be concerned with those regions where there is direct overlap or intersection with the named steganography cyber threat.

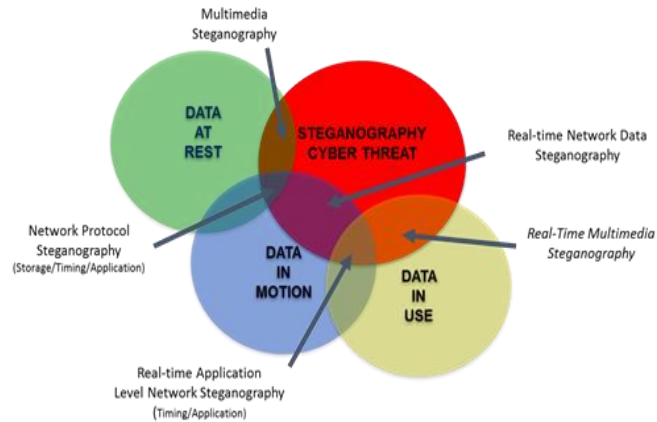

Figure 11. Data Intersects of Steganography Cyber Threats

Looking first at the major intersections of the steganography cyber threat (SCT) domain with the DAR and DIU domains, we define these areas as *multimedia steganography* activities. The subdomain of DAR represents the well-known area of multimedia data hiding. The DIU major subdomain is very similar to the DAR steganography domain in that we define this as *real-time multimedia steganography*. This domain is consistent with the typical data-at-rest form of steganography with the major difference being the operations are performed at the application level, actively by the user. An example of this type of real-time application embedding is the *hCovert* covert channel tool which is implemented using HTTP GET requests between webservers during an active communications session [40]. Embedding tools, such as the *hCovert* application, for this subdomain are not as prolific as the static multimedia steganography domain; however we predict tool capabilities to continue to emerge.

Additional subdomain sets of interest in the Venn diagram are formed by the direct intersections with the data-in-motion (DIM) domain. The resultant intersections define three sub-domains and it is these three regions that are of most interest in this paper. The symbolic definitions for these regions are shown in Table 2 below.

Table 2. Network Steganography Symbolic Definitions

| Class | Type | Abbreviation | Definition |
|---|---|---|---|
| I | Real-Time Network Steganography | RT-NS | DIM ∩ SCT |
| II | Network Protocol Steganography | NPS | DAR ∩ DIM ∩ SCT |
| III | Real-time Application Network Steganography | RT-ANS | DIM ∩ DIU ∩ SCT |
| IV | *Real-time Network Data Steganography* | *RT-NDS* | *(RT-NS)–[NPS+(RT-ANS)]* |

The first minor subdomain of these regions is the classical network protocol steganography, or NPS type of covert communications. One unique clarifying characteristic of NPS methods is that they operate in a transitory state between data-at-rest and data-in-motion. This unique feature is possible only because the end-user system has full control of the formation and management of the datagrams. Recall that NPS implementation occurs in region 2 of the protocol space, before it is actually presented to the lower level network interface handler of protocol region 3. Because of the dependent properties between static and dynamic data transitioning, we choose to omit the real-time moniker from the NPS definition.

A minor subdomain observable in Figure 11 is *real-time application-level network steganography*, or RT-ANS, which is closely related to network protocol steganography, in that it also resides in a transitory state, namely the data-in-use to data-in-motion domains. The major difference between RT-ANS and NPS and is that with RT-ANS, execution of steganography occurs using active applications as opposed to traditional, well-known static toolsets. Specifically, the active applications in RT-ANS are not used to form the region 2 data structures (ref Fig 3). The operating system takes full control of the application presentation to the lower datagram assembly processes in the lower region [12]. Real-time operations are maintained and the dynamic nature of this domain is maintained. Examples of this type of real-time steganographic functionality are the *VoVoIP* toolset or Skype embedding as is described in the *SkyDe* [29]. The *SoCat* application, which opens named pipes between users to establish a covert data channel, is another excellent example from the RT-ANS threat subdomain, exhibiting dynamic covert communications properties [40].

Next, the majority subdomain in Venn diagram is formed by the full intersection of the data-in-motion and the steganography cyber threat. We define this composite region as *real-time network steganography* as indicated in Table 2. The entirety of this region envelopes the concepts of classic network steganography as well as real-time application steganography. We derived our unique definition for real-time network data steganography as shown in Table 2.

Class IV network steganography defines our real-time network data steganography (RT-NDS). This type consists of the discrete region which is characterized by the condition that data-in-motion interacts with the steganography cyber threat fully independent of any end-user systems or application processing before the data is set into motion. Data modifications in this class of network steganography occur strictly while the data is in transit between the network origination and destination end points.

## 4. REAL-TIME NETWORK DATA STEGANOGRAPHY

We further define RT-NDS as a distinct subset of the overarching real-time network steganography domain with unique characteristic operational properties. To illustrate this distinction, we refer to Figure 12 which depicts the data structure for a network Ethernet frame from within the region 3 protocol space. From this figure, we observe the span of effect for each of the various type of steganography. It should be noted that information transitions between networking nodes and remains in a region 3 state until it is delivered to an end destination.

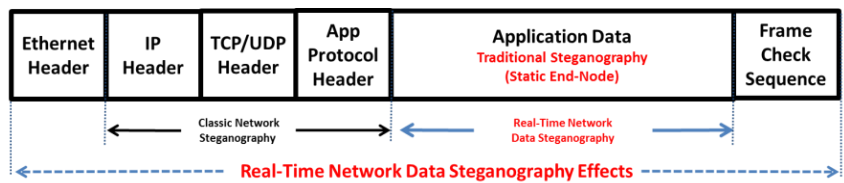

Figure 12. Span of Real-time Network Data Steganography

Recall that with traditional *multimedia steganography*, covert data is embedded in multimedia cover files and transmitted from end-to-end nodes within static data segments. In this instance, the effects on the datagram structure are fixed in region 1 and no user node operation occurs outside of this protocol space. In classical *network steganography*, the datagram headers in the protocol space of region 2 are used to embed covert data. These specific operations still occur at the end-user nodes, manipulated and managed without affecting the application data, again derived from the region 1 space. Thus, we reemphasize the condition that traditional steganography and classical network steganography have bounded and constrained functionality set at the end-user nodes.

### 4.1 REAL-TIME DATA TRANSFER CONSIDERATIONS

*Real-time network data steganography* varies markedly from the previous embedding types. The targeted data object (intended carrier file) in the datagram is the multimedia application data constructed during application data segmentation. Data embedding in RT-NDS operations occurs during the actual transmission phase of the data transfer. This is the time period between when the data is transmitted over the transmission medium by the sender to the time when the data is recovered from the medium by the receiver. No direct action from end-user nodes occurs on the datagrams. As shown in Figure 12 above, the entire data structure is considered during RT-NDS functions, to include the lower level network datalink construct, in this case Ethernet.

Whereas traditional network steganography involves the modification of protocol structures as its main covert channel carrier, real-time network steganography encompasses the entire range of the transmitted data structure with modification of the entire datagram and network datalink level construct during data-in-motion. More to the point, modifications are performed on individual data streams as data transfers over packetized networks. It should be clear then that given the range of data control in RT-NDS, operations in this data domain can involve or even implement classic network steganography algorithms, but in real-time using intermediate nodes.

We now consider an example by showing active data operation actions involving datagrams entering and leaving this RT-NDS subdomain. Figure 13 represents the high-level processes involved in transmitting an image between two network nodes. The image shown consists of a typical 512x512 BMP format with the three-color layers, red, blue and green. With these three layers and the fixed size of the file, the calculated nominal transfer size is 262,144 x 3 layers or 786,432 bytes. File metadata format bytes are also included in the gross file transfer for canonical image data structures. The final size of the file to be transferred in just over 786,460 bytes.

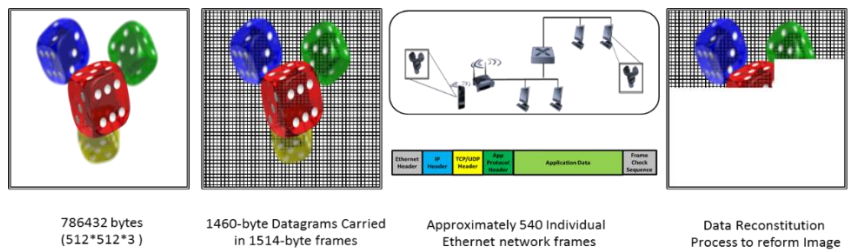

Figure 13. Image Segmentation and Transmission

This figure depicts the image first being segmented by the end-user application and data networking layers to properly allocate the data space available in the individual packets. This segmentation on this size of image will create roughly 540 individual data segments, packaged and presented to the network as 1460-byte datagrams. The addition of the IP and TCP along with the Ethernet header and frame check sequence value will result in roughly 540 individual network frames, each with a size of 1514 bytes.

Under the conditions that this image file is in transit, we have stated that RT-NDS involves the modification of the *application data portion* of each of these Ethernet frames as they are transported across the network. Given that the network frame is already constructed and in the transmission state (data-in-motion), any changes to the application data space will (as shown in Figure 12 above) have direct effects across the entire network frame. For example, the TCP header contains a 16-bit checksum value as part of its 20-byte structure. This checksum is calculated over the application data and the TCP header itself and is used to ensure that no modifications or data errors occur during transmission. Therefore, changes to the application data necessitate the recalculation of the TCP header checksum [12].

The IP header also has a 16-bit CRC checksum field, however this value is only applicable to the IP header itself and does not include the encapsulated TCP or application data it is routing [12]. IP header checksums are calculated by the intermediate router to account for mutable field changes such as the time to live or TTL field, which incidentally changes at each routing node [12]. Additionally, each time an Ethernet frame is generated for transmission, such as between router nodes or on the last network segment before delivery to the receiver node, a frame check sequence, or FCS, is also calculated. This FCS ensures the entire datagram has maintained its integrity between network nodes. Any change in the datagram will therefore require a recalculation of the FCS. Therefore, any changes to the data field of transiting datagrams will necessitate a recalculation of both the TCP CRC value and the Ethernet FCS [12].

Referring once again to the data structure of Figure 12 we note that RT-NDS affects the same application data space as traditional static application steganography. The major difference between these two types is exemplified by the fact that RT-NDS steganography happens outside of and beyond control of the end-user node whereas traditional steganography occurs before the application data is presented to the region 2 network protocol space. Therefore, for traditional steganography, no consideration is given to the TCP, IP or network datalink levels for this type of covert communications.

This is also the case for classical network protocol steganography. Specifically, NPS operations occurs directly at the end nodes before the datagram is presented to the network datalink level. Because the datagram is constructed at the originating end-user nodes, any steganography relevant changes will be incorporated directly in the TCP checksum value which will then not change during end-to-end data delivery. If the IP fields are used for covert communications, as in the case of an IP storage channel, the IP checksum is originally calculated at the point of origin for all immutable fields that are being used to implement the channel and the normal mutable fields, affected by routing will be modified as needed by the routing functions. RT-NDS clearly involves some additional network operational burdens that are not a factor for traditional steganography or classical network protocol data channels.

**4.2 REAL-TIME NETWORK DATA INTERCEPT CONSIDERATIONS**

When transmitting data, such as an image file between end-user nodes, each of the individual network frames will traverse over the network links and, as was previously discussed, the reliable transport, delivery, and presentation of the image data is managed by TCP operational exchanges between the sender and the receiving nodes. We know these individual datagrams are routed and analyzed by any intermediate routing devices to ensure the TCP segments are properly presented to the end system to be accounted for by the receiving application [13].

Moreover, multimedia data files are transferred from servers to clients, or between end-user nodes in non-monolithic fashion. Specifically within networks, data files are transmitted as segmented data structures using datagrams [13]. While TCP monitors and manages the data flow, there is no guarantee that the data will be presented to the receiver in the same order as it was transferred from the sender [12]. A number of network considerations come into play as data flows over and between networks from end-point to end-point. Conditions such as a change of datagram routing due to network congestion may result in packets being delivered out of order [13]. Loss of actual network frames and datagrams, due to intermediating network collisions can also result in the need to retransmit lost data. These are just two of many unpredictable circumstances that affect network data communications [13].

The implications of these network frame intercept considerations, under conditions that likely involve out of order or lost datagrams are significant. The successful application of real-time network data embedding into well-ordered multimedia data segments is critical for proper end-point extraction of the intended covert data. We continue our introduction and exploration of network data steganography by introducing a generalized network intercept model.

### 4.3 REAL-TIME SYSTEM INTEGRATION

We further clarify the concept of real-time network data interception and embedding using the model shown in Figure 14. This figure shows two end point devices that may be transmitting a targeted multimedia file. In this figure, four potential intercept and embedding scenarios are shown by devices that may reside at various points on the network. These notional systems are capable of performing the aforementioned operations of frame interception, analysis, modification, and retransmission necessary to implement real-time network data steganography. These RT-NDS system operate unbeknownst to the end point systems or the intermediate routing network nodes and normal communications is maintained by these systems.

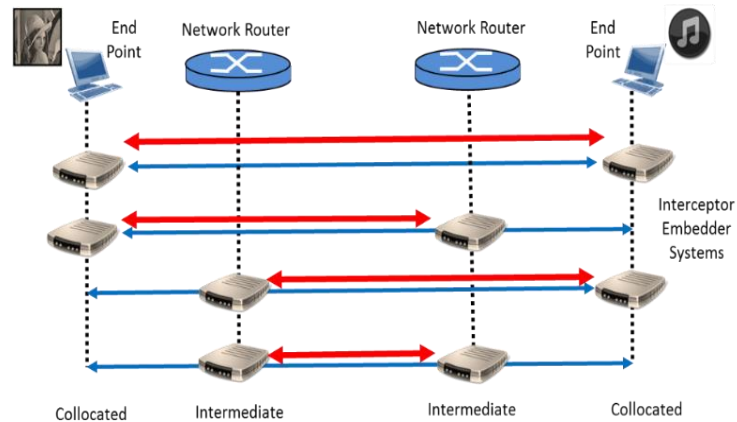

Figure 14. Real-time Network Steganography Access Points

This network configuration with an integrated real-time intercept system is a prime example of the real-time network steganography cyber threat. Under these conditions the means to embed covert data in various multimedia content is conceivable.

### 5. PERFORMANCE ANALYSIS

Table 3 and Figure 15 are six representative image files used to demonstrate the real-time embedding of BMP images by the RT-NDS system. In our initial experiments, the simple LSB +/- method was used [4]. For the RGB image files only the LSB value of the red layer will be changed.

Table 3. Representative Image Parametric Data Set

| Representative Image Set (bmp) | Structure | Dimensions | Size (Kbytes) | LSB Embedding Capacity |
|---|---|---|---|---|
| Peppers | RGB | 512x512x3 | 768 | 262,144 |
| Baboon | RGB | 512x512x3 | 768 | 262,144 |
| Barbara | Grayscale | 510x510x3 | 764 | 260,100 |
| Lena | Grayscale | 512x512x3 | 768 | 262,144 |
| Diamond | Grayscale | 256x256x1 | 65 | 65,536 |
| House | Grayscale | 256x256x1 | 65 | 65,536 |

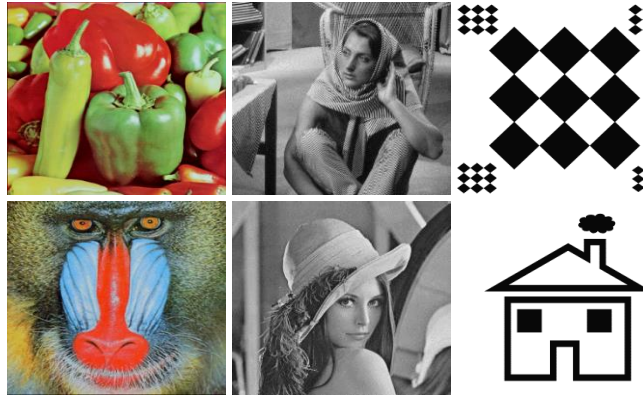

Figure 15. Representative Image Set

## 5.1 Real-Time Results and Analysis

Table 4 shows the results of the representative image transfers over the test network, passing through the RT-NDS system. To demonstrate the effectiveness of the embedding process, 100% of the least significant bits were modified for the targeted red-layer. As shown in the Table 4 the total transmission time difference for the files ranges from an instantaneous delay of 35 milliseconds for the 65 Kb file size to approximately 575 milliseconds of delay for the larger 768 Kb image files. Essentially the impact of the RS-NDS is about ½ a second delay for an average sized image file.

Table 4. Real-Time Steganography System Performance

| Representative Image Set | Size (Kbytes) | Transmission Speed (ms) | | Difference (ms) |
|---|---|---|---|---|
| | | Direct | RT-NDS | |
| Peppers | 768 | 85.67 | 661.01 | 575.35 |
| Baboon | 768 | 83.15 | 642.12 | 558.97 |
| Barbara | 764 | 81.18 | 531.23 | 450.05 |
| Lena | 768 | 83.48 | 627.15 | 543.67 |
| Diamond | 65 | 7.28 | 42.82 | 35.54 |
| House | 65 | 8.59 | 43.23 | 34.64 |

Greater analysis shows that the overall network degradation is approximately 13%. Given that the system is modifying every third byte in the real-time data stream over the entire length of the file with an overall average delay of less than ½ second, this performance should be considered acceptable. Additionally, given the preliminary testing was performed on commodity hardware with non-optimized prototype software indicates many paths for improved performance. Efficiency gains can be expected with an optimized code base running on dedicated processors or specialized programmable devices (e.g. DSP or FGPA).

## 6. CONCLUSIONS

In this paper, we have introduced new concepts and implementation methodologies relating to a novel capability called *real-time network data steganography*. There is a lack of published research for this specific domain, therefore some key contributions for this emerging technology include:

- Defines symbolic representation of technology domain in relation to cyber security threats
- Introduces novel network steganography techniques using real-time stream analysis
- Describes non-disruptive modification methods of multimedia files in real-time data streams

This research explores real-time network data steganography where we describe how this new embedding method compares and contrasts with traditional notions of network steganography or static data application-based steganography. We also discussed some unique real-time network operations considerations for embedding data in internetworked data streams. Finally this basic research document is intended to form the groundwork for future research in the domain of real-time network data embedding technologies.

**Authors**

**James C. Collins** received a B.S. degree in electrical engineering from Arizona State University, Tempe Arizona in 1987 and a M.S. degree in electrical engineering from Southern Methodist University, Dallas Texas in 1999. He is currently pursuing his Ph.D. degree in electrical engineering, specializing in digital signal processing theory at the University of Texas at San Antonio, Texas. His research interests include embedded systems, network systems security, data hiding, and image processing.

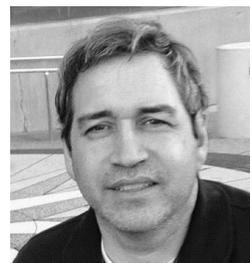

**Sos S. Agaian** is a Peter T. Flawn Professor of Electrical and Computer Engineering at the University of Texas, San Antonio (UTSA), and Professor at the University of Texas Health Science Center, San Antonio. He received a M.S. degree (summa cum laude) in mathematics and mechanics from Yerevan University, Armenia, and a Ph.D. degree in math and physics from the Steklov Institute of Mathematics, Russian Academy of Sciences, as well as a PhD degree in Engineering Sciences from the Institute of the Control System, Russian Academy of Sciences. He has authored more than 500 scientific papers, 7 books, and holds 14 patents. Some of his major research efforts are focused in multimedia processing, imaging systems, information security, artificial intelligence, computer vision, 3D imaging sensors, image fusion, and biomedical and health Informatics.

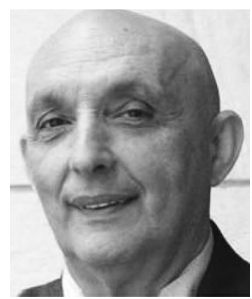